\documentclass[twocolumn,tightenlines,superscriptaddress,showpacs]{revtex4-2}

\usepackage{amsmath}
\usepackage{amssymb,amsfonts,latexsym}
\usepackage{bm}
\usepackage[mathcal]{euscript}
\usepackage{graphicx}
\usepackage{epsfig}

\usepackage[svgnames]{xcolor}
\usepackage{xfrac}
\usepackage{mathrsfs}

\usepackage[colorlinks,linkcolor=MediumBlue,citecolor=MediumBlue,urlcolor = MediumBlue]{hyperref}
\usepackage{url}

\allowdisplaybreaks

\graphicspath{{Figures/}}

\newcommand{\kT}{k_{\rm B}T_{\rm eff}}
\newcommand{\Teff}{T_{\rm eff}}
\newcommand{\Tc}{T_{\rm c}}
\newcommand{\Te}{T_{\rm e}}
\newcommand{\bJ}{\bar{J}}
\newcommand{\rmd}{{\rm d}}
\newcommand{\br}{\bar{\rho}}
\newcommand{\rc}{\rho_{\rm c}}
\newcommand{\Vb}{V_{\rm b}}
\newcommand{\Ms}{M_{\rm s}}
\newcommand{\Fc}{F_{\rm c}}
\newcommand{\phic}{\phi_{\rm c}}
\newcommand{\betac}{\beta_{\rm c}}

\begin{document}

\title{Reentrant condensation transition in a model of driven scalar active matter with diffusivity edge}

\author{Jonas Berx}
\affiliation{Max Planck Institute for Dynamics and Self-Organization (MPIDS), 37077 G\"ottingen, Germany}
\affiliation{Institute for Theoretical Physics, KU Leuven, B-3001 Leuven, Belgium}

\author{Aritra Bose}
\affiliation{Max Planck Institute for Dynamics and Self-Organization (MPIDS), 37077 G\"ottingen, Germany}

\author{Ramin Golestanian}\email{ramin.golestanian@ds.mpg.de}
\affiliation{Max Planck Institute for Dynamics and Self-Organization (MPIDS), 37077 G\"ottingen, Germany}
\affiliation{Rudolf Peierls Centre for Theoretical Physics, University of Oxford, Oxford OX1 3PU, United Kingdom}

\author{Beno\^{\i}t Mahault}\email{benoit.mahault@ds.mpg.de}
\affiliation{Max Planck Institute for Dynamics and Self-Organization (MPIDS), 37077 G\"ottingen, Germany}

\date{\today}

\begin{abstract}
The effect of a diffusivity edge is studied in a system of scalar active matter confined by a periodic potential and driven by an externally applied force. We find that this system shows qualitatively distinct stationary regimes depending on the amplitude of the driving force with respect to the potential barrier. For small driving, the diffusivity edge induces a transition to a condensed phase analogous to the Bose–Einstein-like condensation reported for the nondriven case, which is characterized by a density-independent steady state current. Conversely, large external forces lead to a qualitatively different phase diagram since in this case condensation is only possible beyond a given density threshold, while the associated transition at higher densities is found to be reentrant.
\end{abstract}

\maketitle

\section*{Introduction}

Systems for which detailed balance is broken at the microscopic scale are known as active matter~\cite{Gompper_2020}.
This property endows them with the ability to exhibit collective behaviour impossible at equilibrium.
Notable examples are the possibility of active particles to phase separate without attractive interactions~\cite{cates_MIPS_2015,Cotton2022,Golestanian_2019},
or to exhibit long-range orientational order in two dimensions~\cite{TT_1995,chate_2020,Mahault_2021_PRL}.

In recent years, intensive efforts have been made to study the rich phenomenology of active systems via minimal models that capture their key features.
Active particles are indeed routinely modelled as persistent walkers~\cite{ABP_review,Howse2007,BechingerRMP2016,Zottl_2016}, and in a lattice gas setting as agents with an internal polarity setting a preferred hopping direction~\cite{Thompson_2011,Solon_2013_PRL,Soto_2014_PRE,Kourbane_2018_PRL}.
Without large-scale orientational order, the macroscopic dynamics of active systems is often captured by a nonlinear 
drift-diffusion equation describing the evolution of the particle density field~\cite{Chavanis2008,cates_MIPS_2015,Kourbane_2018_PRL,Chakraborty2020}. 
Furthermore, because such equation describes a dynamics evolving far from equilibrium, 
its effective diffusivity and mobility {\it a priori} do not satisfy a Fluctuation Dissipation Relation (FDR)~\cite{Kubo_FDT_1966}.

Due to the interplay between activity and particle interactions, 
some mean field active dynamics may be formulated in terms of an effective diffusivity that vanishes or becomes negative beyond a certain density threshold~\cite{cates_MIPS_2015,Meng_2021}.   
The class corresponding to a diffusivity remaining identically zero at large enough densities, which hereafter we will refer to as \emph{diffusivity edge}, was introduced phenomenologically in~\cite{Golestanian_2019}.
There, it was shown that the coupling of the diffusivity edge with harmonic confinement triggers the formation of a point-like condensate at the ground state of the potential,
while the corresponding transition exhibits remarkable similarities with Bose-Einstein Condensation (BEC)~\cite{ZIFF1977}.
More recently, this BEC-like condensation transition was reported in numerical simulations of a model of magnetic microswimmers confined in a quasi-one-dimensional channel~\cite{Meng_2021}
The BEC-like transition observed in the diffusivity edge class is furthermore reminiscent of condensation phenomena arising in various mass transport models~\cite{Evans2005,Evans2006,Evans_2014_JPhysA,Majumdar1998,Majumdar2000,Chakraborty2021}.

\begin{figure*}[t]
     \centering
     \includegraphics[width=\linewidth]{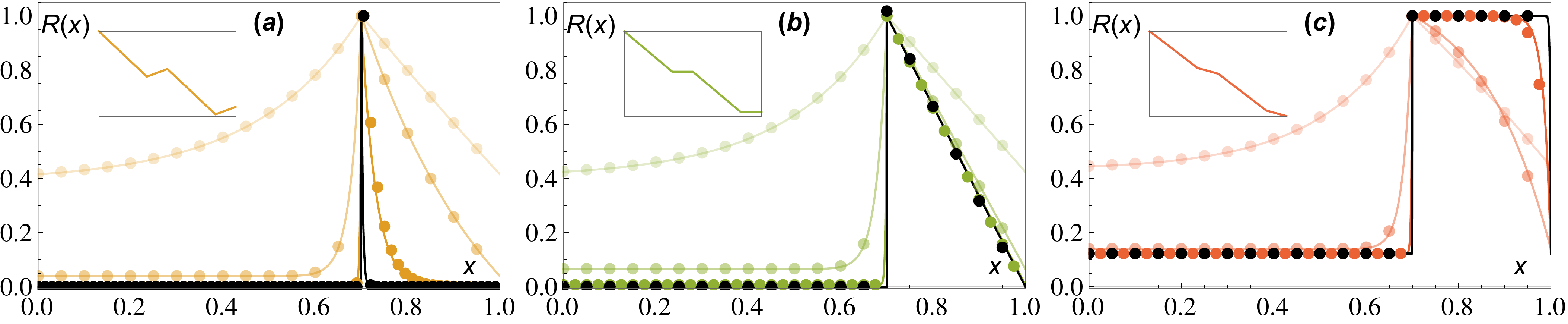}
     \caption{Stationary density profiles~\eqref{eq:R} for $\alpha = \tfrac{7}{10}$ in the three regimes described in the text: panels (a,b,c) respectively correspond to $f = -0.7$, $f=0$ and $f = 0.2$.
     Curves with increasing transparency correspond to an increase of the effective temperature $\Teff$. In all panels dots indicate numerical simulation results, while solid lines correspond to the theoretical prediction. The insets sketch the profile of the potential $U(x)$ corresponding to each case.
     }
     \label{fig_density_row}
 \end{figure*}

In this work, we study the influence of an external driving on the behaviour of a scalar active matter system presenting a diffusivity edge.
Working in one dimension, we achieve a nonvanishing steady state current by confining the dynamics in a tilted periodic potential.
This setup plays a fundamental role in the modelling of a plethora of physical systems~\cite{Risken} including molecular motors~\cite{JulicherRMP1997}, enzymatic cycles~\cite{AgudoPRL2021}, 
electrophoresis of polyelectrolytes~\cite{Ajdari1991,Nixon1996}, rotating dipoles in external fields \cite{Reguare2000}, Josephson junctions~\cite{Barone1982}
and active ratchets~\cite{Angelani_2011EPL,ReichhardtAnnrev2017}.
Although for weak external driving forces such that the potential has local minima the phenomenology of condensation resembles that observed at zero drive~\cite{Mahault_2020},
the strong force regime where the potential is monotonous leads to striking differences.
For strong forces our study indeed reveals the existence of a particle density threshold below which condensation cannot arise.
The condensation transition at large enough densities, moreover, leads to the formation of a condensate occupying a finite volume, 
such that upon cooling the system may undergo a subsequent evaporation transition, leading to reentrance.
All the theoretical results presented below are systematically compared to direct numerical simulations of the continuous theory governing the dynamics (details about numerical methods are given in~\cite{SM}).


\section*{Scalar active matter with diffusivity edge} \label{sec:model}

We consider the following one-dimensional phenomenological evolution equation for a scalar density field $\rho(x,t)$:
\begin{equation}
    \label{eq:continuity}
     \partial_t \rho + \partial_x {J} = 0\,, \qquad  {J}= - M(\rho)\rho\partial_x U - D(\rho)\partial_x\rho\,,
\end{equation}
where $U({x})$ is an externally applied potential, while $M(\rho)$, and $D(\rho)$ respectively denote the $\rho$-dependent effective mobility and diffusivity. 
When the microscopic dynamics satisfies detailed balance, the ratio between $M(\rho)$, and $D(\rho)$ obeys a FDR and is thus equal to the temperature of the system.
With broken detailed balance, on the contrary, $D(\rho)/M(\rho)$ is not constrained by any FDR.
A number of non-interacting active systems, however, 
satisfy a generalised FDR accounting for the renormalization of their effective diffusivity by activity~\cite{ABP_review,cates_MIPS_2015,Meng_2021,Davide2008PRE}.
Here, we therefore assume such a relation in the dilute limit, and define an effective temperature as
\begin{equation}
    \label{eq:tuning}
    \kT \equiv \lim_{\rho \to 0} 
    \frac{D(\rho)}{M(\rho)} ,
\end{equation}
which we will moreover use as a control parameter for the dynamics.
For finite densities, due to the nonequilibrium character of active dynamics the ratio $D(\rho)/M(\rho)$ has {\it a priori} no reason to remain equal to $\kT$.
In particular, here we impose a diffusivity edge beyond a threshold $\rc$ such that 
$D(\rho)/M(\rho) = 0$ for $\rho \ge \rc$.

When $U(x)$ is harmonic, such diffusivity edge induces the formation of a 
singular condensate at the ground state for $\Teff$ below a transition temperature $\Tc$~\cite{Golestanian_2019}.
Remarkably, although the dynamics described by eq.~\eqref{eq:continuity} is fully classical the corresponding condensation transition carries signatures of BEC.
The BEC-like condensation moreover holds for periodic potentials presenting degenerate minima~\cite{Mahault_2020}.
To investigate how the BEC-like condensation transition is modified by the presence of an external drive, we consider here the external potential 
$U(x) = V(x) - F x$,
whose periodic part $V(x)$ has a peak value $\Vb$ at $x=0$ and satisfies $V(x+L) = V(x)$ with $L$ denoting the period,
while $F$ is a stationary uniform driving force.


\section*{The steady state solution} \label{sec:sol}

We now calculate the steady state solution of eq.~\eqref{eq:continuity}, which satisfies $\partial_x J = 0$.
For $F \ne 0$, it is thus associated with a non-vanishing uniform current $\bJ$.
We moreover assume a constant mobility $M(\rho) = \Ms$, and a step diffusivity profile such that
\begin{equation} \label{eq_D_step}
D(\rho) =  \Ms \kT\,\Theta(\rc -\rho)\,,
\end{equation}
where $\Theta$ is the Heaviside step function. 
This simple way to implement the diffusivity edge, while it allows for analytical progress, 
does not qualitatively modify the properties of the condensation transition at $F = 0$ as compared to more realistic mobility and diffusivity profiles~\cite{Golestanian_2019,Mahault_2020}.
As we discuss in the final section, 
for $F \ne 0$ the properties of the system  are also largely insensitive to the specific shape of $D(\rho)$.
In what follows, we rescale space and time such that $L$ and $\Ms \Vb$ are set to unity. 

Assuming that the dynamics evolves on a ring, due to the translational invariance of the problem 
we express stationary density profiles over a single period of $V(x)$.
With the above choice of parameters, we derive the steady state solution of eq.~\eqref{eq:continuity} via usual techniques~\cite{Risken} (see also the Supplementary Material~\cite{SM} for calculation details). 
Defining $\beta \equiv (\kT)^{-1}$ as the inverse effective temperature and assuming the density to be lower than $\rc$ everywhere, we obtain
\begin{equation}
    \label{eq:density_rho<rhoc}
    \rho(x) = \beta \Vb \bJ e^{-\beta U(x)} \left(\frac{I_+(1)}{1-e^{-\beta F}} - I_+(x)\right)\,,
\end{equation}
with $I_+(x) \equiv \int_0^x \exp\left[\beta U(x')\right]\mathrm{d}x'$.
The expression of the steady state current is then obtained from the normalization condition $\br = \int_0^1 \rmd x \, \rho(x)$, with $\br$ denoting the mean density in the system~\cite{SM}.

Although the integrals in the expression of the steady state profile~\eqref{eq:density_rho<rhoc} can be numerically determined for an arbitrary periodic potential $V(x)$, 
it is instructive to consider the case of a sawtooth potential with anisotropy parameter $\alpha \in (0;1)$, such that
\begin{equation}
\label{eq:potential}
    V(x) = \Vb\left(1 - \frac{x}{\alpha}\right) \times 
    	\begin{cases}
       1 & 0 \le x \le \alpha\\
        \alpha/(\alpha-1) & \alpha \le x \le 1 
    \end{cases} ,
\end{equation}
and which is to be extended periodically with unit period. 
Note that with this choice the potential $U(x)$ is symmetric under the combined transformations $F\rightarrow-F$ and $\alpha\rightarrow1-\alpha$, 
such that we only consider positive values of $F$ without loss of generality.
The piecewise linearity of~\eqref{eq:potential} conveniently leads to an exact expression for the density profile~\eqref{eq:density_rho<rhoc}.
Denoting $\rho_0$ as the maximum value of the density, we find that for $\rho_0 < \rc$ the expression for the full profile can be recast as $\rho(x) = \rho_0 R(x)$ 
with
\begin{equation} \label{eq:R}
R(x) \! = \! \frac{1}{\gamma} \! \begin{cases} \left(1 - e^{\beta \Vb f}\right) z^{\tfrac{x}{\alpha}} \! - \! \alpha f \left(ze^{\beta \Vb f}-1\right) \!\!  & 0 \le x \le \alpha  \\
 \left( z - 1\right) e^{\beta \Vb f \tfrac{x - \alpha}{1 - \alpha}} & \\
 \qquad\qquad- (1 + \alpha f) \left(ze^{\beta \Vb f} - 1\right) & \alpha \le x \le 1
\end{cases} ,
\end{equation}
where we have defined the dimensionless force $f \equiv F / \Fc - 1$ with $\Fc \equiv \Vb/ (1-\alpha)$, and
\begin{equation*}
z \equiv e^{\tfrac{\beta \Vb (1 + \alpha f)}{1-\alpha}}, \quad \gamma \equiv z \left( 1  - e^{\beta \Vb f}  \right) - \alpha f \left(ze^{\beta \Vb f}  - 1\right).
\end{equation*}
$\rho_0$ can moreover be expressed in terms of the steady state current $\bJ$, which yields
 \begin{equation}
     \label{eq:maximal_density}
     \rho_{0} = \frac{\bJ (1-\alpha)}{f (1 + \alpha f)} \left(\alpha f + \frac{z(1-e^{\beta \Vb f})}{1-z e^{\beta \Vb f}}\right)\,.
 \end{equation}
 
Looking back at the expression of the potential~\eqref{eq:potential}, it is clear that $\Fc$ corresponds to the value of the external forcing such that the full potential $U$ becomes monotonous. 
In the following, we thus distinguish between the dynamical regimes associated with $f<0$ ($F < \Fc$) and $f>0$ ($F > \Fc$), which we will respectively refer to as subcritical and supercritical.
As shown in fig.~\ref{fig_density_row}, the solution~\eqref{eq:R} in the subcritical regime consists of two convex branches decaying from $\rho_0$ at $x = \alpha$.
For $f = 0$, on the contrary, due to the vanishing potential slope in the range $\alpha \le x \le 1$ the corresponding branch decays linearly from $\rho_0$,
while in the supercritical regime it takes a concave shape.
The maximum value of the density, $\rho_0$, typically increases with $\beta \Vb$ and $\br$, while it decreases with increasing $f$.
In the limit $\beta \Vb \gg 1$ of small effective temperatures the expression of $\rho_0$ moreover simplifies as
\begin{equation} \label{eq:rho0_lims}
\rho_0 \underset{\beta \Vb \gg 1}{\sim} \frac{\br}{1-\alpha} \begin{cases}
-\beta \Vb f(1 + \alpha f) & (f<0) \\
2 & (f=0) \\
\tfrac{(1-\alpha)(1 + \alpha f)}{1 + \alpha (f-1)} & (f > 0)
\end{cases} ,
\end{equation}
while for large effective temperatures fluctuations dominate over the confinement, such that the density profile is nearly uniform and $\rho_0 \sim \br$ 
for $\beta \Vb \ll 1$ and for all $f$.

The solution~\eqref{eq:R}, however, holds only so long as $\rho_0$ remains below the diffusivity edge $\rc$. 
When the condition $\rho_0 = \rc$ is satisfied, the stationary density profile indeed becomes singular at $x = \alpha$ which reflects the
fact that the system undergoes a condensation transition~\cite{Golestanian_2019}.
We now analyse the properties of this transition and the nature of the condensate and discuss how they are affected by the presence of a nonzero driving force.
From now on we fix $\alpha = \tfrac{7}{10}$ for all numerical evaluations.
In all figures theoretical results are displayed with solid lines while numerical data from simulations of eq.~\eqref{eq:continuity} are shown with symbols. 


\begin{figure}[t]
     \centering
     \includegraphics[width=0.98\columnwidth]{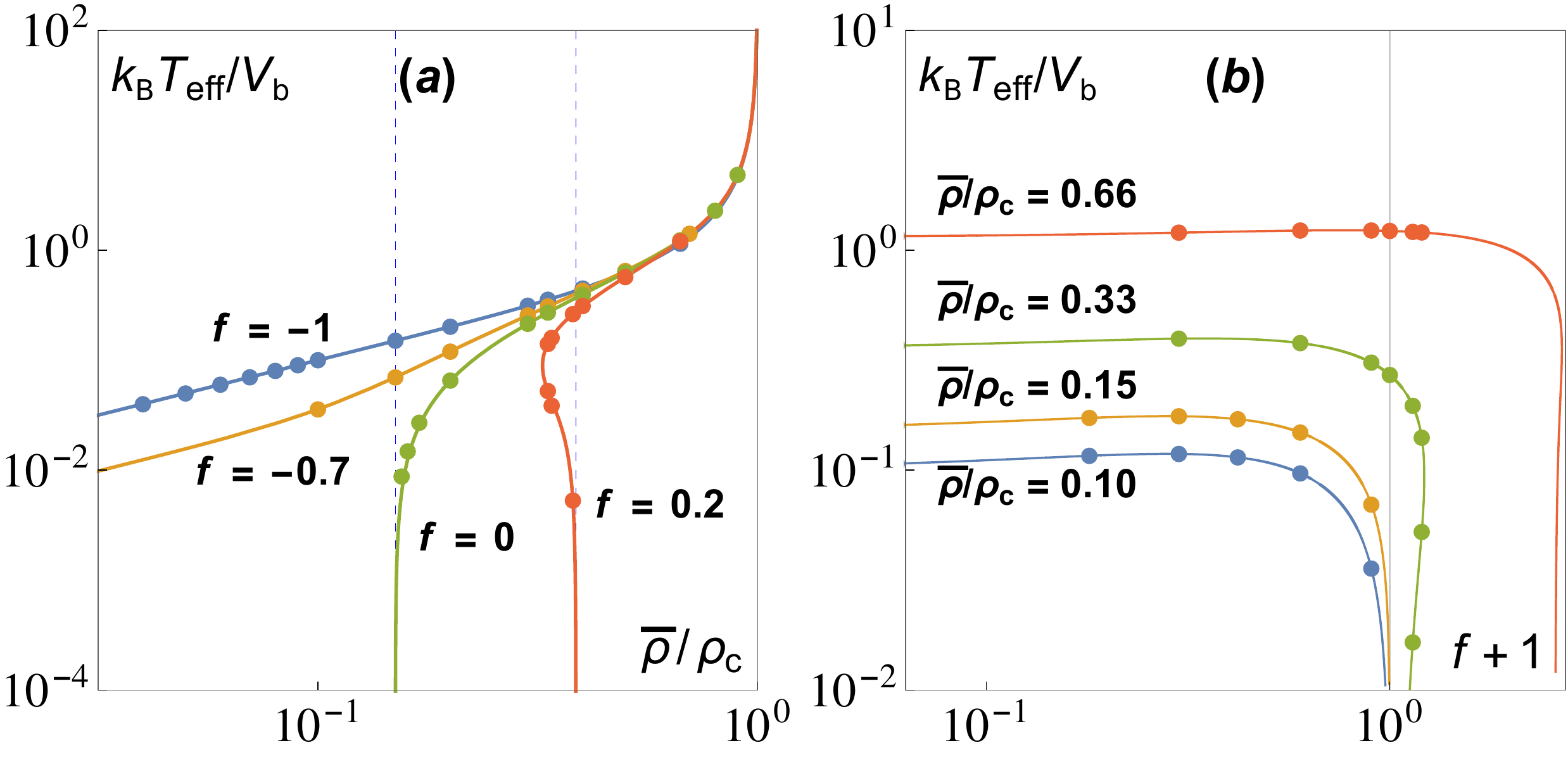}
     \caption{Phase diagrams computed from the condition $\rho_0 = \rc$, showing the effective transition temperature as a function of mean density $\br$ at fixed $f$ (a) 
     or as function of $f$ at fixed $\br$ (b).
     In panel (a) the two vertical dashed lines mark the density ratio $\lambda$ (cfr. eq.~\eqref{eq:lambda_cases}) below which no condensation occurs in the limit $\Teff \to 0$.
    }
     \label{fig_phase_diagrams}
 \end{figure}

\section*{The subcritical regime}
For $f < 0$ the potential $U(x)$ exhibits a local minimum at $x = \alpha$. This case is therefore qualitatively similar to the nondriven case addressed in ref.~\cite{Mahault_2020}.
Setting $\rho_0 = \rc$ in eq.~\eqref{eq:maximal_density} gives the formal condition on the parameters $\beta \Vb$, $f$ and $\br$ for the condensation to occur.
The resulting transition lines at fixed $f$ and $\br$ are shown respectively in fig.~\ref{fig_phase_diagrams}(a,b) (see the blue and yellow curves).
They predict the existence of a transition for all $\br < \rc$, while the transition temperature in the limit of vanishing densities reads
\begin{equation} \label{eq:Tc_small_F}
\frac{k_{\rm B}\Tc}{\Vb} \underset{\br \ll \rc}{\sim} -\frac{\br}{\rc} \frac{f(1+\alpha f)}{1-\alpha}.
\end{equation}
 
\begin{figure}[tp]
     \centering
     \includegraphics[width=0.98\columnwidth]{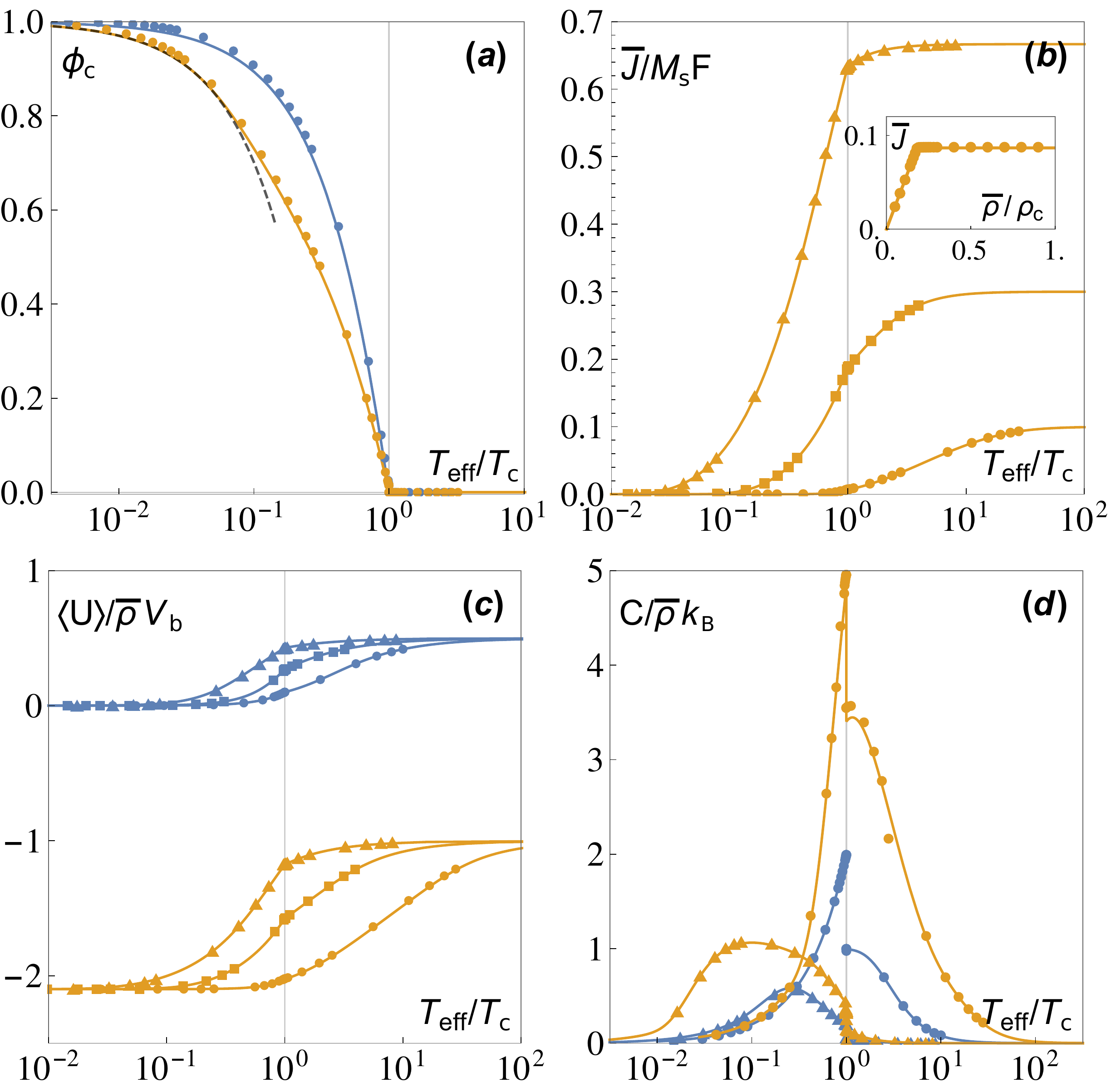}
     \caption{Subcritical ($f < 0$) transition to condensation with $f=-1$ (blue) and $f=-0.7$ (orange). 
     (a) The condensate fraction $\phic$  for $\bar{\rho}/\rc = 0.33$ (the black dashed line shows the low temperature approximation~\eqref{eq_low_large_T_approx}), 
     (b,c,d) The stationary current $\bJ$(b), mean internal energy(c) and heat capacity(d) as function of the effective temperature. The inset in (b) shows $\bJ$ vs. $\br$ at fixed $\Teff=0.1$, condensation occurs at $\br/\rc \approx 0.19$.
     Dots, squares and triangles respectively correspond to $\br/\rc = 0.1$, $0.3$ and $0.66$.}
     \label{fig_transition_sub}
 \end{figure}

For $\Teff < \Tc$, the density profile in the condensed phase can moreover formally be written as
\begin{equation} \label{eq:rho_cond_subcrit}
	\rho(x) = \br \phic \delta(x - \alpha) + \rc R(x) \qquad (\Teff < \Tc),
\end{equation}
with the function $R(x)$ defined in eq.~\eqref{eq:R}.
This solution thus consists of a smooth part with maximum density given by $\rc$ and a singular condensate located at the local potential minimum $x = \alpha$.
The condensate fraction, $\phic$, is then determined from the density normalization. 
The expression of $\phic$ in the general case is rather cumbersome and does not yield much physical insight, but taking the limits of small and large effective temperatures it simplifies as 
\begin{equation} \label{eq_low_large_T_approx}
\!\! \phic \underset{\beta \Vb \gg 1}{\sim} 1 - \frac{\Teff}{\Tc} , \quad 
\phic \underset{\beta \Vb \ll 1}{\sim} \left(1-\frac{\rc}{\bar{\rho}}\right)\left(1-\frac{\Tc}{T_{\rm eff}}\right).
\end{equation}
These expressions correspond to those derived in~\cite{Mahault_2020}, and match well with the full solution in the low temperature regime (fig.~\ref{fig_transition_sub}(a)). 
The presence of a weak driving force thus does not qualitatively 
modify the scaling of the condensate fraction with effective temperature but only affects the value of the transition temperature $\Tc$.

Since the condensate is located at a local minimum of the potential, it does not contribute to the global current $\bJ$.
Therefore, the latter is obtained from eq.~\eqref{eq:maximal_density} by replacing $\rho_0$ with $\rc$, namely
\begin{equation} \label{eq:J_subcritcond}
\bJ = \frac{\rc f (1+\alpha f)}{(1-\alpha)} \left(\alpha f + \frac{z(1-e^{\beta \Vb f})}{1-z e^{\beta \Vb f}}\right)^{-1} \, (\Teff < \Tc).
\end{equation}
Remarkably, $\bJ$ does not depend on the mean particle density $\br$, 
which reflects the fact that increasing $\br$ does not affect the smooth part of the distribution~\eqref{eq:rho_cond_subcrit} (cfr. inset of fig.~\ref{fig_transition_sub}(b)), but only leads to an increase of the condensate fraction $\phic$.
This feature is analogous to the divergence of the isothermal compressibility at $\Teff = \Tc$ found in the nondriven case~\cite{Golestanian_2019,Mahault_2020}, 
which implies that the confining pressure exerted by the potential is independent of $\br$.

The BEC-like condensation transition at $F=0$ is moreover associated with a singular behaviour of several observables such as 
the mean internal energy $\langle U \rangle \equiv \int_0^1 \rmd x \, \rho(x)U(x)$ or the heat capacity $C \equiv \rmd \langle U \rangle / \rmd \Teff$~\cite{Golestanian_2019}.
For $F>0$, the explicit expressions of these thermodynamic functions, although they can straightforwardly be obtained from eq.~\eqref{eq:R}, are generally quite lengthy and we do not report them here. Some asymptotic results are given in the Supplementary Material~\cite{SM}.
Figure~\ref{fig_transition_sub}(c,d) shows that their behavior at the transition is qualitatively similar to the nondriven case, with $\langle U \rangle$ exhibiting a discontinuous slope at $\Teff = \Tc$, resulting in a discontinuous jump of $C$.
In addition, in the limit of strong confinement, the mean particle current can be expressed as
\begin{equation} \label{eq:J_sub_lowT}
\bJ \! \underset{\beta \Vb \gg 1}{\sim} \! -\frac{\rc f (1+\alpha f)}{1-\alpha} e^{-\tfrac{1-\alpha}{1+\alpha f}\tfrac{\rc}{\br} \tfrac{\Tc}{\Teff}} 
\begin{cases}
\tfrac{\Tc}{\Teff} & (\Teff \ge \Tc) \\ 
1 & (\Teff < \Tc)
\end{cases},
\end{equation}
which highlights that the condensation transition is associated with a cusp in the decay of $\bJ$ with the effective temperature (fig.~\ref{fig_transition_sub}(b)).
Finally, eq.~\eqref{eq:J_sub_lowT} also indicates that the presence of a condensate leads to a faster decay of the current with the effective temperature.



\section*{The supercritical regime}
We now address the case $f\ge 0$, which marks a qualitative change in the phase diagram of the system (see the green and red curves in fig.~\ref{fig_phase_diagrams}).
Indeed, it appears from eq.~\eqref{eq:rho0_lims} that for $f \ge 0$ the condensed phase can be reached at zero effective temperature only for $\br \ge \lambda \rc$
with
\begin{equation}
\label{eq:lambda_cases}
\lambda \equiv \begin{cases} 
(1-\alpha)/2 & (f=0) \\
(1 - \alpha +\alpha f)/(1+\alpha f) & (f > 0)
\end{cases}.
\end{equation} 

\begin{figure}[t]
     \centering
     \includegraphics[width=.98\columnwidth]{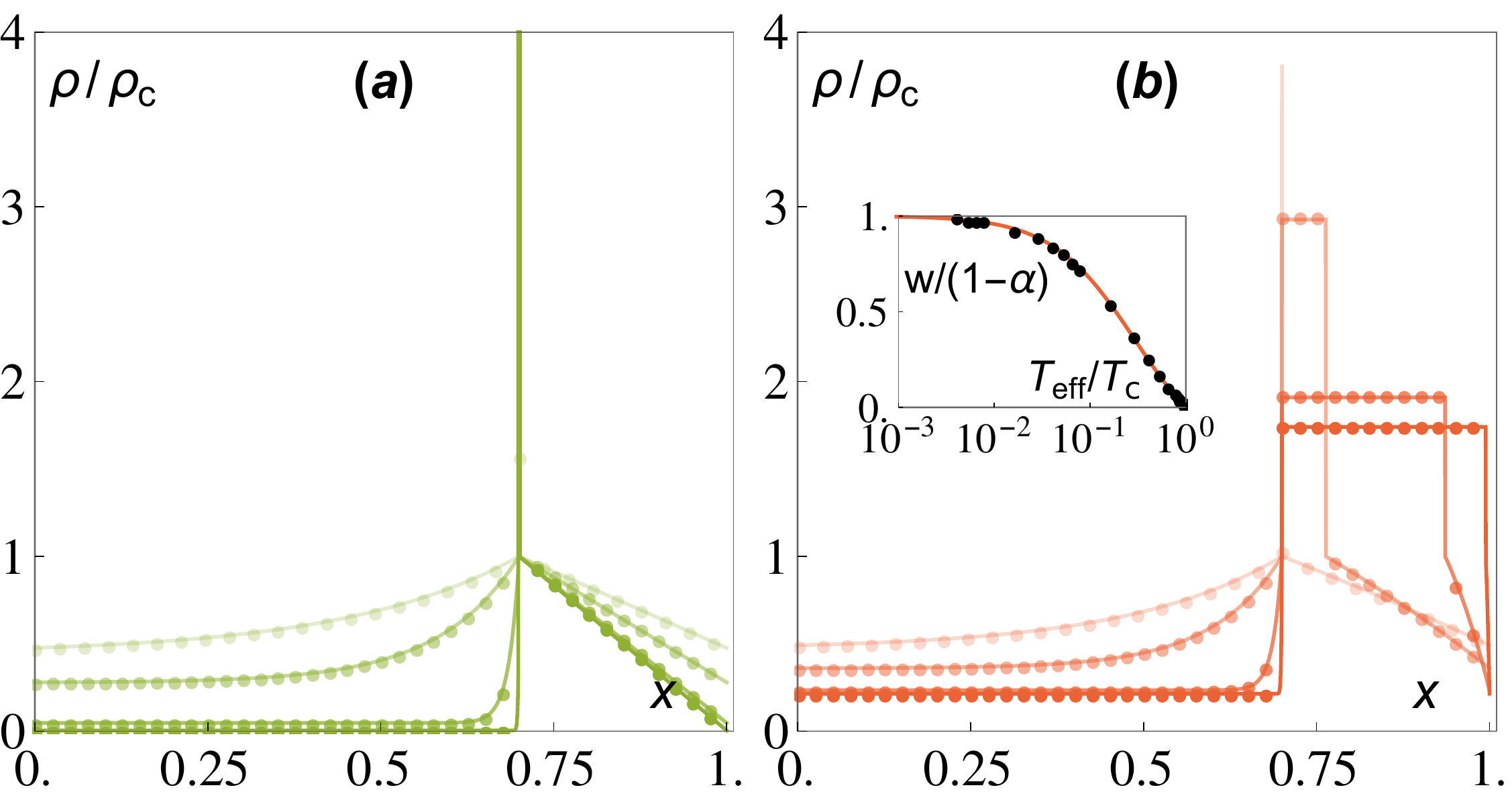}
     \caption{Condensed density profiles for $f = 0$(a) and $f = 0.2$(b) at fixed $\br/\rc = 2/3$ for various effective temperatures.
     Curves with increasing opacity correspond to decreasing $\Teff$.
     In (a) the point-like condensate density is truncated for clarity.
     The inset in (b) shows how the condensate width varies with $\Teff$.
     }
     \label{fig_condensed_profiles_super}
 \end{figure}

Taking $f=0$, numerical simulations of eq.~\eqref{eq:continuity} reveal that the condensate remains point-like (fig.~\ref{fig_condensed_profiles_super}(a)).
Therefore, density profiles in the condensed phase are described by eq.~\eqref{eq:rho_cond_subcrit}.
This case is thus analogous to the subcritical regime, with however some qualitative differences in the low effective temperature limit.
Taking  $f=0$ and $\beta \Vb \gg 1$ in eq.~\eqref{eq:J_subcritcond}, we indeed find that the current vanishes linearly with the effective temperature:
\begin{equation*}
\bJ \underset{\Teff \to 0}{\sim} \frac{\rc}{1-\alpha}\frac{\kT}{\Vb} \qquad (\Teff < \Tc,\, f=0),
\end{equation*}
in contrast with the faster exponential decay found in the subcritical regime (see eq.~\eqref{eq:J_sub_lowT}).
Correspondingly, we find that in the limit of zero effective temperature the condensate does not contain all the mass of the system but carries a finite fraction equal to
$\phic \underset{\Teff \to 0}{\sim} 1 - \lambda \rc/\br$.
This feature can be explained by the fact that, since for $f=0$ the potential is no more confining, 
it allows for spatial coexistence between a point-wise condensate at $x=\alpha$ and a residual gas phase even for $\Teff = 0$.
Figure~\ref{fig_condensed_profiles_super}(a) shows that this gas phase is indeed located in the region where the net force balances to zero, 
and hence can persist even for $\Teff = 0$ without inducing a global current.

Considering now the supercritical regime at $f > 0$, we find from numerical simulations of eq.~\eqref{eq:continuity} that the condensate is only point-like at $\Teff = \Tc$, 
while for $\Teff < \Tc$ it consists of a uniform domain of finite density extending from $x=\alpha$ with a width $w$ (fig.~\ref{fig_condensed_profiles_super}(b)).
Therefore, $f=0$ marks a depinning transition from a point-wise condensate to an extended condensate occupying a finite volume.
Consequently, for $f > 0$ cooling down the system from $\Tc$ may result in the condensate density becoming lower than $\rc$ at $\Teff \equiv \Te < \Tc$, 
leading to a low-temperature evaporation transition.
We find that such a scenario typically occurs for $\br \lesssim \lambda \rc$, and therefore that in the supercritical regime the transition to condensation is reentrant,
as shown in the phase diagrams of fig.~\ref{fig_phase_diagrams}.
In~\cite{SM} we moreover provide a theoretical argument for the supercritical reentrant transition in the low effective temperature limit.
Reentrant condensation transitions have been observed in models of two-species zero-range processes~\cite{DAGA2017}, 
while in the context of active matter reentrance was also predicted for motility induced phase separation~\cite{Paoluzzi2020}.


To study the condensed phase at $f > 0$, we describe the density profile as a piecewise function defined on the three intervals $[0;\alpha]$, $[\alpha;x^*]$ and $[x^*;1]$ with $x^* \equiv \alpha + w$.
Solving eq.~\eqref{eq:continuity} for a steady state with constant current $\bJ$ for $x \le \alpha$ and $x^* \le x$, while assuming a uniform condensate of density $\br \phic/w > \rc$ in between, we obtain
\begin{equation} \label{eq:sol_rho_supercritcond}
\rho(x)  \underset{f > 0}{=}  \begin{cases} \frac{\bJ \alpha (1-\alpha)}{1+\alpha f}\left(1 - z^{\tfrac{x}{\alpha}-1} \right) \! + \! \rc z^{\tfrac{x}{\alpha}-1}  & 0 \le x \le \alpha  \\
\br \phic/w & \alpha < x <  x^* \\ 
 \frac{\bJ(1-\alpha)}{f}\left(1 - e^{\beta \Vb f\tfrac{x-x^*}{1-\alpha}} \right) & \\
\qquad\qquad\qquad + \rc e^{\beta \Vb f\tfrac{x-x^*}{1-\alpha} } & x^* \le x \le 1
\end{cases},
\end{equation}
where we have used the fact that $\rho(\alpha) = \rho(x^*) = \rc$.
The periodicity condition $\rho(0) = \rho(1)$ then allows to express the condensate width $w$ as function of the current $\bJ$.
Finally, a closed expression for the current is obtained from $\bJ = -\Vb^{-1} \int_0^1 \rmd x \, U'(x)\rho(x)$ which,
after eliminating $\phic$,
leads to (details in~\cite{SM})
\begin{align} \label{eq:J_super_cond}
\bJ \underset{f>0}{=} & \frac{\rc (1+\alpha f)}{\kappa} \left[1-\frac{1}{z} + \frac{\beta \Vb f (1+\alpha f)}{1-\alpha} \frac{\br}{\rc}\right] \; (\Teff < \Tc),
\end{align}
where $\kappa \equiv \alpha(1-\alpha)(1-z^{-1}) + \lambda \beta \Vb (1+\alpha f)^2$.

Since for $f>0$ the condensate occupies a finite volume, it contributes to the global current that now explicitly depends on the mean particle density, 
in contrast with eq.~\eqref{eq:J_subcritcond} derived in the subcritical regime.
Equation \eqref{eq:J_super_cond} can then be used to calculate the condensate width $w$ as well as the condensate fraction $\phic$~\cite{SM}.
Close to the transition, we find that both expressions simplify as
\begin{equation*}
w \underset{\Teff \lesssim \Tc}{\propto} \frac{\Tc}{\Teff} - 1 , \qquad 
\phic \underset{\Teff \lesssim \Tc}{\propto} 1 - \frac{\Teff}{\Tc}  \qquad (f > 0),
\end{equation*}
such that near the threshold the condensate density decays linearly with the effective temperature: $\br \phic/w \propto \Teff/\Tc$.

\begin{figure}[t]
     \centering
     \includegraphics[width=0.98\columnwidth]{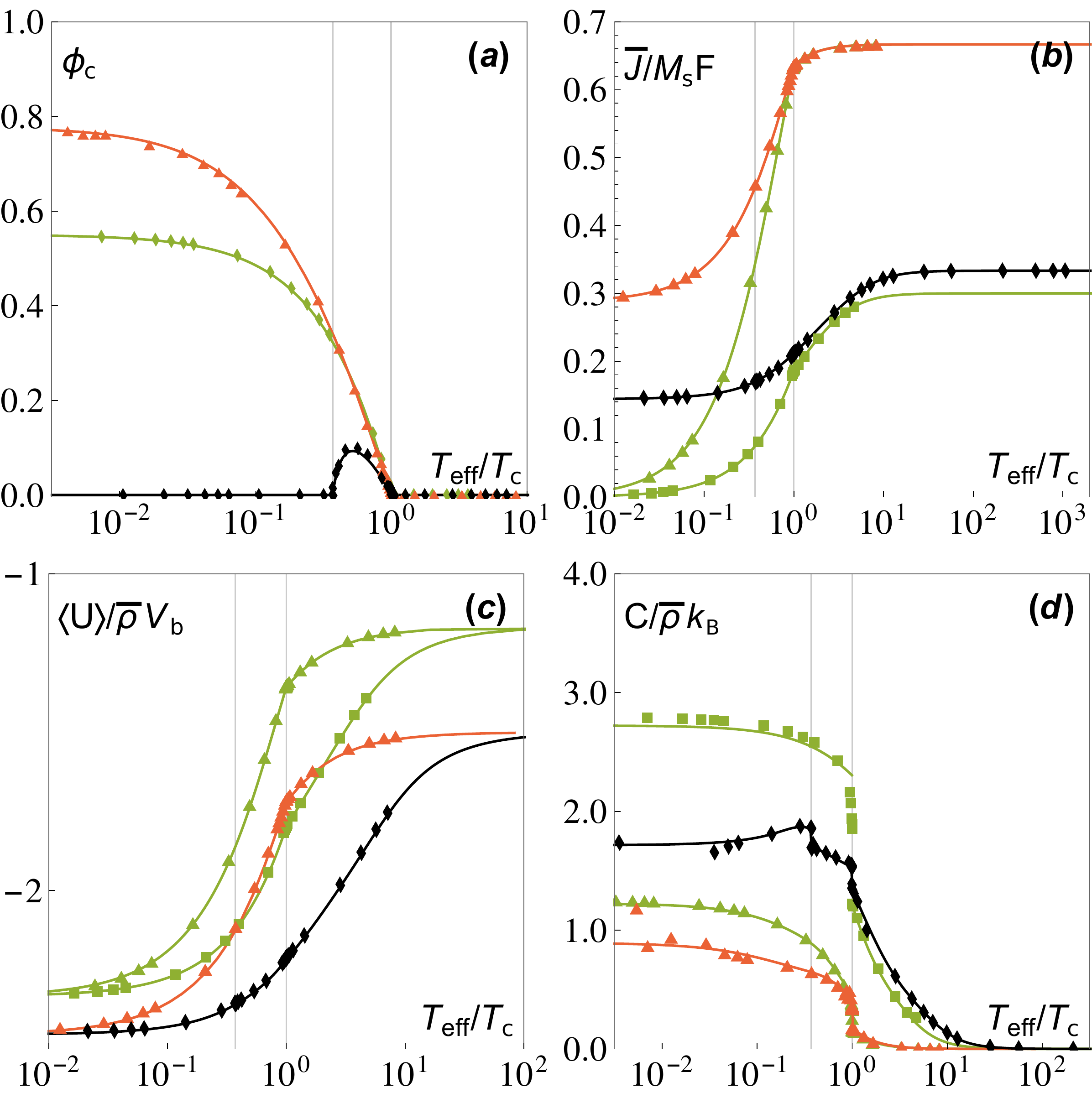}
     \caption{Transition to condensation in the supercritical regime ($f \ge 0$), with $f=0$ (green)  and $f=0.2$ (red and black). Squares, diamonds and triangles respectively correspond to $\br/\rc =0.3$, $0.33$ and $0.66$. Vertical lines indicate the condensation and evaporation (happening only for the data shown in black) thresholds $\Tc$ and $\Te$.} 
     \label{fig_thermo_super}
 \end{figure}

In the condensed phase, we moreover find that the solution~\eqref{eq:sol_rho_supercritcond} 
matches perfectly with the profiles obtained from numerical simulations of eq.~\eqref{eq:continuity} (fig.~\ref{fig_condensed_profiles_super}(b)). 
Furthermore, the thermodynamics of condensation carries the BEC signatures similarly to the subcritical regime (fig.~\ref{fig_thermo_super}).
The current and mean energy indeed exhibit a discontinuous slope at $\Teff = \Tc$ associated with a discontinuous jump of the heat capacity.
Remarkably, in the reentrant regime these features also hold for the evaporation transition at $\Teff = \Te$, despite the condensate not being point-like in this case.

In the limit of small effective temperatures, we find that the condensate width approaches $1-\alpha$ from below as
\begin{equation*}
1-\alpha - w \underset{\Teff \to 0}{\propto} \frac{\Teff}{\Tc} \qquad (f>0),
\end{equation*}
while the current and condensate fraction converge to
\begin{equation*}
\bJ \underset{\Teff \to 0}{\sim} \frac{\br f}{(1-\alpha)\lambda}, \quad \phic \underset{\Teff \to 0}{\sim} 1 - \frac{\alpha^2 f}{(1+\alpha f)\lambda}, \quad (f>0).
\end{equation*}
As a consequence of the depinning transition, in the supercritical regime the current saturates to finite values for $\Teff = 0$.
Similarly to the $f=0$ case, the condensate fraction moreover saturates to a finite value, which highlights a gas-condensate coexistence.
Indeed, the density distribution at $\Teff = 0$ and $f>0$ simply consists of two disconnected plateaus with respective values
\begin{equation*}
\rho(x) \underset{\Teff \to 0}{\sim} \frac{\br}{\lambda} \begin{cases}
\alpha f/(1+\alpha f) & 0 \le x \le \alpha \\
1 & \alpha < x \le 1 
\end{cases} \qquad (f > 0).
\end{equation*}


\section*{Generalization to the sinusoidal potential}

\begin{figure}[tp]
     \centering
     \includegraphics[width=\columnwidth]{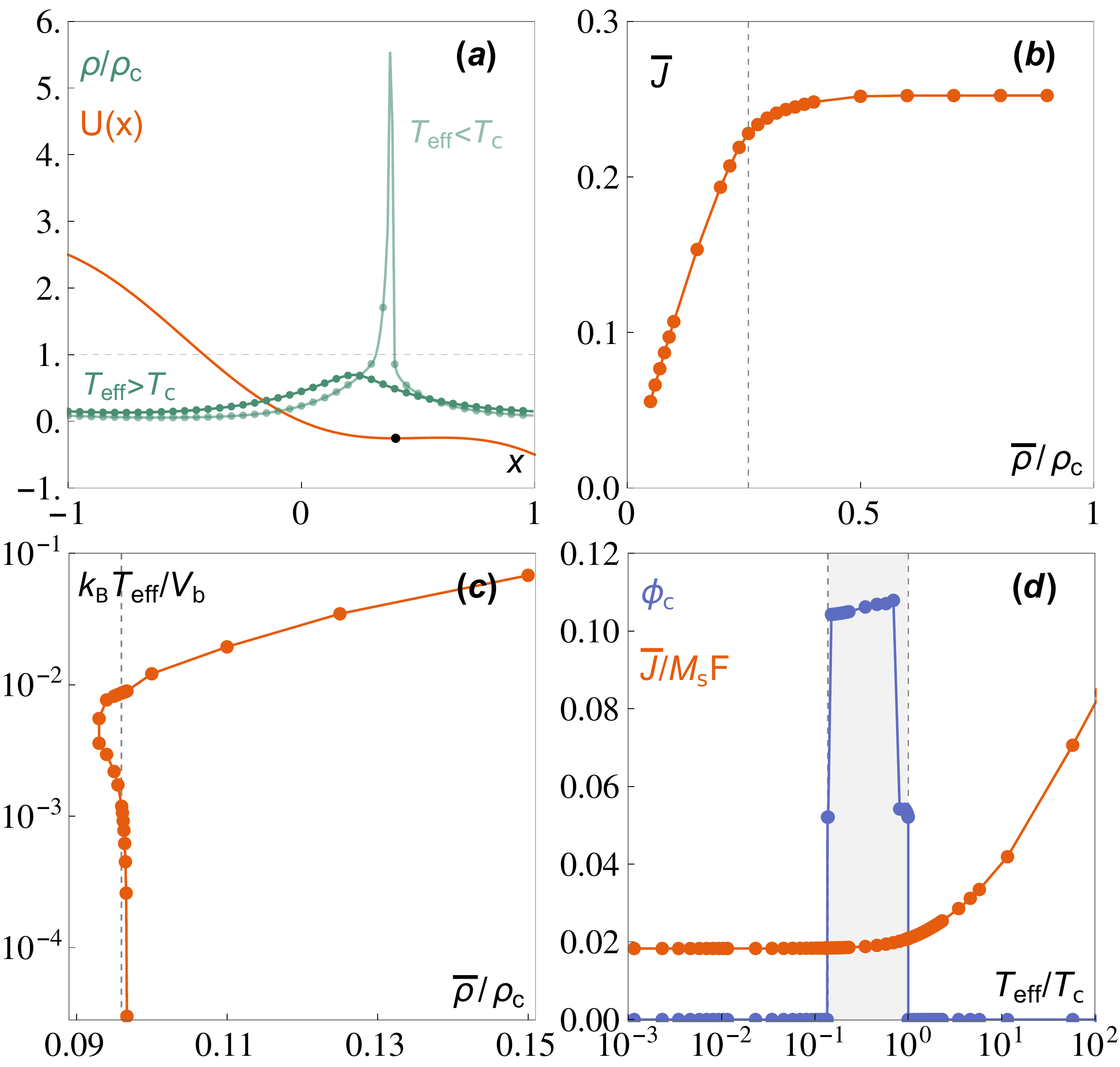}
     \caption{Condensation transition in a sinusoidal potential with $\Vb =1$. (a) Density profiles for $\Teff > \Tc$ and $\Teff < \Tc$ with $f=-0.05$. The potential is also shown with its minimum (black dot).
     (b) Scaling of $\bJ$ with the mean density while crossing the transition at $\br/\rc \approx 0.26$ at fixed $\Teff = 0.5$.
     (c) Phase diagram showing $\Tc$ as a function of $\br$.
     (d) Current and condensate fraction as a function of $\Teff$ at $\br/\rc \approx 0.096$ and $f \approx 0.02$.
     Here lines are drawn as a guide for the eye.}
     \label{fig_sine}
 \end{figure}

All the results obtained so far correspond to the step diffusivity profile~\eqref{eq_D_step} and the sawtooth periodic potential~\eqref{eq:potential}, which allow for exact derivations. 
To assess their generality, we now consider a different setting for which the diffusivity edge is reached as (keeping the previous notations)
\begin{equation} \label{eq_D_quad}
D(\rho) =  \Ms \kT (1-\rho/\rc)^2\, \Theta(\rc -\rho)\,,
\end{equation}
while the periodic potential is sinusoidal: $V(x) = \tfrac{\Vb}{2} \left[ 1 - \cos(\pi x)\right]$.
As expected, this choice leads to density profiles smoother than obtained previously with the sawtooth potential (fig.~\ref{fig_sine}(a)).
Looking at the condition for the full potential to be monotonous, the threshold associated with the depinning transition is in this case $\Fc = \pi \Vb/2$.

In the subcritical regime for which $f = F/\Fc - 1 \le 0$, numerical simulations of eq.~(\ref{eq:continuity}) 
reveal that the phenomenology of the transition is similar to that described previously.
Given a mean particle density $\br < \rc$ and a driving force $f     < 0$, 
there always exists a value $\Tc > 0$ of $\Teff$ such that the maximum density reaches $\rc$ and the system condenses.
In this case, however, the maximum of the density profile~\eqref{eq:density_rho<rhoc} is distinct from the local minimum of the total potential $U(x)$ (fig.~\ref{fig_sine}(a)).
Therefore, the condensate is generally not point-like but has a finite width.
Consequently, the condensate contributes to the current $\bJ$ which increases sub-linearly with the mean particle density for $\Teff < \Tc$ (fig.~\ref{fig_sine}(b)).
The condensate width, however, is in general fairly small and converges to zero as $\Teff \to 0$, 
such that its effect on $\bJ$ is hardly appreciable in numerical simulations. 

Taking now $f > 0$, similarly to what was described previously for the sawtooth potential there exists a finite density threshold below which no condensation occurs (fig.~\ref{fig_sine}(c)).
Noting that the supercritical regime is defined as when the potential becomes strictly monotonous,
the zero-temperature density profile indeed obeys in steady state (assuming that $U$ is smooth)
$\bJ = -\rho(x) U'(x)$ such that, after using the normalization condition
\begin{equation}
\rho(x) \underset{f > 0}{=} \br\,\left[U'(x) \int_0^1 \frac{\rmd y}{ U'(y)} \right]^{-1} \qquad (\Teff = 0).
\end{equation}
Hence, we deduce that in the limit of vanishing effective temperatures condensation only happens for $\br \ge \lambda \rc$ with
\begin{equation}
\lambda = \min_x[|U'(x)|] \int_0^1 \frac{\rmd y}{|U'(y)|} .
\end{equation}
For the sinusoidal potential, we find $\lambda = \sqrt{f/(2+f)}$ which is in agreement with our numerical simulations (fig.~\ref{fig_sine}(c)).
More remarkably, our numerical results point towards a reentrant transition with the system undergoing an evaporation transition at $\Teff = \Te < \Tc$, highlighting the generality of this feature
(figs.~\ref{fig_sine}(c,d)).


\section*{Conclusion}

We have studied the effect of nonzero steady state currents on the BEC-like condensation induced by a diffusivity edge.
In contrast with the nondriven case, the condensate at nonzero drive might occupy a finite volume
similar to what was observed in mass transport models with finite-range interactions~\cite{Evans2006,Evans_2014_JPhysA}.
Moreover, the stationary current is found to be essentially independent of the mean particle density in the condensed phase of subcritical regime,
a feature that has also been described for the externally driven zero-range process~\cite{Schutz_2007}.  
The most prominent differences with respect to previous studies of the diffusivity edge class were observed in the supercritical regime.
There, we found that condensation can only occur beyond a minimum particle density, 
while the corresponding transition is reentrant at moderate densities due to the presence of an evaporation transition at low effective temperatures. 

The results presented here highlight the similarities between the diffusivity edge class and a variety of mass transport models. 
Characterising further these similarities at the dynamical level, \textit{e.g.}\ to study the coarsening of the condensate or
to investigate the presence of kinematic waves such as those observed in the zero-range process~\cite{Schutz_2007}, requires to go beyond the mean field level considered here.
Introducing noise in the dynamics would moreover allow to study the interplay between other features of externally driven Brownian motion,
such the enhanced effective diffusion~\cite{Reimann2001,Reimann2002,Reimann_2002Report}, with the BEC-like condensation.

\acknowledgments
This work has received support from the Max Planck School Matter to Life and the MaxSynBio Consortium, which are jointly funded by the Federal Ministry of Education and Research (BMBF) of Germany, and the Max Planck Society.

\bibliography{biblio.bib}

\pagebreak

\onecolumngrid

\setcounter{equation}{0}
\setcounter{figure}{0}
\renewcommand{\theequation}{S\arabic{equation}}
\renewcommand{\thefigure}{S\arabic{figure}}

\begin{center}
\textbf{\Large Supplementary Material}
\end{center}

\section{Derivation of the density profile in presence of a tilted potential for constant diffusivity and mobility}\label{app:A}

Here we provide calculation details on the derivation of the density profile for a constant mobility $\Ms$ and diffusivity $D = \Ms\kT$.
In the dimensionless units defined in the main text the corresponding density current reads 
\begin{equation} \label{eq_SM_J}
J = -\frac{\kT}{\Vb} \,\rho(x) \,\frac{\rmd[\ln(\rho(x)) + \beta U(x)]}{\rmd x} ,
\end{equation}
where the potential follows $U(x) = V(x) - F x$ with $V(x)$ a periodic function with unit period and peak value $\Vb$ which we assume to be reached at $x = 0$ and $1$.
Assuming a constant current $J=\bJ$ and using the relation $\varphi'(x) = e^{-\varphi(x)}\left(e^{\varphi(x)}\right)'$, we integrate Eq.~\eqref{eq_SM_J} and get for $0 \le x \le 1$,
\begin{equation}
    \label{eq:app_density_derivation_2}
    \rho(x) = e^{-\beta U(x)}\left[\rho_- e^{\beta \Vb} - \beta \Vb \bJ\int_0^x\mathrm{d}x'e^{\beta U(x')}\right]\,,
\end{equation}
where $\rho^* \equiv \rho(x=0)$. As the periodicity of $V(x)$ implies that $U(x + k) = U(x) - k F$ for integer $k$, the density profile for $x > 1$ can be expressed as
\begin{equation}
    \label{eq:app_density_derivation_3}
    \begin{split}
        \rho(x + k) &= e^{\beta [kF - U(x)]}\left[\rho^* e^{\beta \Vb} - \beta \Vb \bJ\int_0^{x + k}\mathrm{d}x' e^{\beta U(x')}\right]\\
        &= e^{\beta [kF - U(x)]}\left[\rho^* e^{\beta \Vb} - \beta \Vb \bJ\left(\int_{k}^{x + k}\mathrm{d}x' e^{\beta U(x')} + \sum_{p=0}^{k-1}\int_{p }^{(p+1)}\mathrm{d}x' e^{\beta U(x')}\right)\right]\\
        &= e^{\beta[kF - U(x)]}\left[\rho^* e^{\beta \Vb} - \beta \Vb \bJ\left(I_{+}(x)e^{-\beta k F} + I_{+}(1)\sum_{p=0}^{k-1}e^{-\beta p F}\right)\right]\,,\\
    \end{split}
\end{equation}
where we have defined $I_\pm(x) \equiv \int_0^x\mathrm{d}x'e^{\pm \beta U(x')}$. After computing the geometric sum, the density profile satisfies
\begin{equation}
    \label{eq:app_density_derivation_4}
    \begin{split}
        \rho(x + k) &=e^{-\beta U(x)}\left[e^{\beta k F}\left(\rho^* e^{\beta \Vb} - \beta \Vb \bJ\frac{I_{+}(1)}{1-e^{-\beta F}}\right) - \beta \Vb \bJ\left(I_{+}(x) - \frac{I_{+}(1)}{1-e^{-\beta F}}\right)\right]\,.\\
    \end{split}
\end{equation}
Assuming now $F>0$, we eliminate the divergence of $\rho(x)$ for $k\rightarrow\infty$ (or equivalently use the translational invariance of the problem), by imposing the the relation
$(1-e^{-\beta F})\rho^* e^{\beta \Vb} = \beta \Vb \bJ I_{+}(1)$, leading to the relation presented in the main text:
\begin{equation}
    \label{eq:SM_density_rho<rhoc}
    \rho(x) = \beta \Vb \bJ e^{-\beta U(x)} \left(\frac{I_+(1)}{1-e^{-\beta F}} - I_+(x)\right)\,.
\end{equation}

As stated in the main text, the expression of the steady state current is then simply obtained from the condition $\br = \int_0^1 \rmd x\, \rho(x)$.
Namely, its expression in the general case reads
\begin{equation}
    \label{eq:Jss}
    \bJ = \frac{\br \kT}{\Vb} \left(\frac{I_+(1)I_-(1)}{1-e^{-\beta F}} - \int_0^1 \rmd x' e^{-\beta U(x')} I_+(x')\right)^{-1}\,.
\end{equation}

\section{Condensation with a piecewise linear periodic potential}

Here we provide additional calculation details about the condensation transition in presence of a sawtooth periodic potential 
\begin{equation}
\label{eq:SM_potential}
    V(x) = \Vb\left(1 - \frac{x}{\alpha}\right) \times 
    	\begin{cases}
      1 & 0 \le x \le \alpha\\
        \tfrac{\alpha}{\alpha-1} & \alpha \le x \le 1 
    \end{cases} ,
\end{equation}
where $0 < \alpha < 1$ denotes the anisotropy parameter.
As is done in the main text, we define $\Fc \equiv \Vb/(1-\alpha)$ as the value of $F$ beyond which the total potential $U$ is monotonous, 
while $f \equiv F/\Fc - 1$.

\subsection{Transition to condensation and reentrance}

\begin{figure}[htp]
    \centering
    \includegraphics[width=0.6\linewidth]{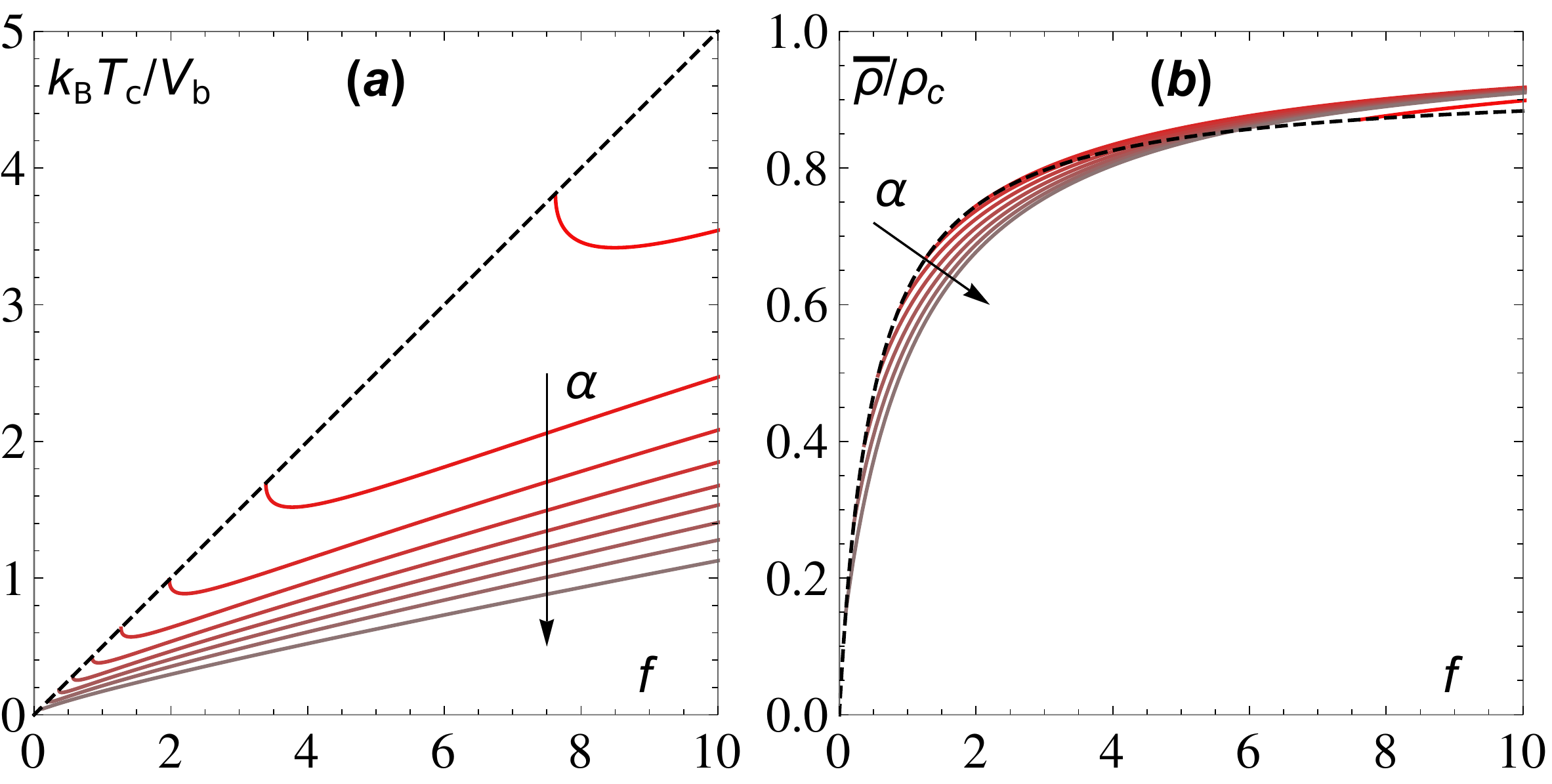}
    \caption{The effective temperature (a) and density ratio (b) of the tip of the reentrance region for $f>0$. The color gradient from red to gray corresponds to increasing values of $\alpha$. Dashed lines indicate the values of $f$ where the temperature and density ratio start to take on real values \eqref{eq:reentrance_tip_real}.}
    \label{fig:reentrance_density_Tc}
\end{figure}

For effective temperatures $\Teff$ above the transition threshold $\Tc$, we find after some calculation that the maximum density $\rho_0$ is given by
\begin{equation} \label{eq_rho0}
\rho_0 = \br \frac{ z (e^{\beta \Vb f}-1) + \alpha f  (z e^{\beta \Vb f}-1)}{(z e^{\beta \Vb f}-1)(1+\alpha (f-1)) - \tfrac{ (1-\alpha)(z-1)(e^{\beta \Vb f}-1) }{\beta \Vb f (1+\alpha f)}},
\end{equation}
where $z \equiv \exp[\beta \Vb (1+\alpha f)/(1-\alpha)]$.
As explained in the main text, the transition temperature $\Tc$ is defined from the condition $\rho_0 = \rc$, with $\rc$ marking the diffusivity edge.
An analytical expression for $\Tc(\br/\rc,f)$ cannot be obtained in general.
However, considering the limit $|\beta \Vb f| \gg 1$ of small effective temperatures we find that the transition threshold condition can be approximated as
\begin{equation}
\frac{\br}{\rc} = \begin{cases}
-\tfrac{1-\alpha}{\betac \Vb f (1+\alpha f)} - (1 -\alpha + \alpha f)e^{\betac \Vb f} + {\cal O}\left((\betac \Vb f)^{-1}e^{\betac \Vb f}\right) & f < 0  \\
 \lambda -\tfrac{1-\alpha}{\betac \Vb f (1+\alpha f)^2} + \tfrac{\lambda}{(1+\alpha f)}e^{-\betac \Vb f} + {\cal O}\left((\betac \Vb f)^{-1}e^{-\betac \Vb f}\right) & f > 0
\end{cases} ,
\end{equation}
where $\lambda \equiv (1 -\alpha + \alpha f)/(1+\alpha f)$ and we have used the notation $\betac \equiv (k_{\rm B} \Tc)^{-1}$.
Keeping only the leading order contributions of the above equations, we recover the expressions given in Eq.~(8) of the main text.
On the other hand, a sufficient condition for the existence of a reentrant transition is that there exists a finite value $\betac \Vb$ such that $\rmd (\br/\rc)/\rmd\beta \Vb = 0$.
After some algebra, we find that the corresponding value of the effective temperature solves
\begin{equation} \label{eq_cond_reentrance}
-\frac{\betac \Vb |f|}{2}e^{-\tfrac{\betac \Vb |f|}{2}} = -\frac{\sqrt{K}}{2}, \quad {\rm with} 
\quad K \equiv \frac{1-\alpha}{1-\alpha + \alpha f} \begin{cases} (1+\alpha f)^{-1} & f< 0 \\
1 & f > 0
\end{cases}.
\end{equation}
Moreover, Eq.~\eqref{eq_cond_reentrance} admits real solutions if and only if $K \le 4 / e^2 \approx 0.54$.
It is easily verified that $K(f < 0) > 1$, such that we do not expect any reentrance in that case.
On the other hand, since $K(f > 0)$ decays to zero with increasing $f$, the supercritical regime is reentrant, at least for $f$ large enough, in agreement with our numerical simulations.
In this case, there are two solutions to Eq.~\eqref{eq_cond_reentrance} which correspond to the two branches of the Lambert $W$ function. 
Considering the branch leading to lower effective temperatures, we finally find that at fixed $f$ the tip of the reentrant region is located in the density effective temperature plane at 
\begin{equation} \label{eq_reentrance_tip}
\frac{\br}{\lambda \rc} = 1 - \frac{K}{(1+\alpha f)\betac \Vb f}\left( 1- \frac{1}{\betac \Vb f} \right), \qquad \betac \Vb f = -2 W_{-1}\left(-\frac{\sqrt{K}}{2}\right) \qquad (f > 0).
\end{equation}
The derivation of Eq.~\eqref{eq_reentrance_tip} only formally holds when $|\betac \Vb f| \gg 1$, such that we do not expect it to hold
too close to the depinning transition at $f = 0$.
Indeed, taking $f \to 0$ eq.~\eqref{eq_reentrance_tip} predicts for instance that the tip of the reentrant region is found at mean particle densities larger than $\lambda \rc$,
which corresponds to the condensation threshold in the limit of vanishing effective temperatures, see fig. \ref{fig:reentrance_density_Tc}. 
The lower threshold for which the effective temperature and density \eqref{eq_reentrance_tip} take on real values can be found to correspond to solving $K=4/e^2$ for $\alpha$, so, $\alpha = (e^2 -4)/(e^2 -4 +4f)$ and subsequently plugging this into \eqref{eq_reentrance_tip},
\begin{equation}
    \label{eq:reentrance_tip_real}
    \frac{\br}{\rc} = f\frac{4-5e^2 -4f+(f+1)e^4}{\left((f+1) e^2-4\right)^2}\,,\qquad \betac \Vb = \frac{2}{f}\,.
\end{equation}
These limiting values are shown as functions of $f$ by dashed black lines in fig. \ref{fig:reentrance_density_Tc}.

\subsection{Density profile in the supercritical condensation regime}

As we discuss in the main text, for $f > 0$ and $\Teff < \Tc$ the system develops a condensate with a finite width $w$
leading the piecewise density profile
\begin{equation} \label{eq:sol_rho_supercriT_cond}
\rho(x)  \underset{f > 0}{=}  \begin{cases} \frac{\bJ \alpha (1-\alpha)}{1+\alpha f}\left(1 - z^{\tfrac{x}{\alpha}-1} \right) \! + \! \rc z^{\tfrac{x}{\alpha}-1}  & 0 \le x \le \alpha  \\
\br \phic/w & \alpha < x <  x^* \\
 \frac{\bJ(1-\alpha)}{f}\left(1 - e^{\beta \Vb f\tfrac{x-x^*}{1-\alpha}} \right) + \rc e^{\beta \Vb f\tfrac{x-x^*}{1-\alpha} } & x^* \le x \le 1
\end{cases} \quad (\Teff < \Tc),
\end{equation}
where $x^* = w + \alpha$ marks the upper limit of existence of the condensate and $\phic$ is the condensate fraction.
Since the potential is linear, the stationary current can simply be obtained from its integral definition as
\begin{align}
\bJ & = -\frac{1}{\Vb} \int_0^1 \rmd x \, U'(x)\rho(x) = \frac{1}{1-\alpha} \left[ \left(\frac{1}{\alpha} + f\right) \int_0^\alpha \rmd x \, \rho(x) + f \br \phic + f \int_{x{^*}}^1 \rmd x \, \rho(x) \right], \nonumber \\
& =  \frac{1}{1-\alpha} \left[ \frac{1}{\alpha} \int_0^\alpha \rmd x \, \rho(x) + f \br \right],
\end{align}
where the last equality was obtained substituting $\phic$ using the normalization condition on $\rho(x)$.
Replacing $\rho(x)$ by its expression given in~\eqref{eq:sol_rho_supercriT_cond} and solving for $\bJ$ we finally obtain
\begin{equation}
\bJ \underset{f>0}{=} \frac{\rc (1+\alpha f)}{1-\alpha} \frac{(1-\alpha)(z-1) + z\beta \Vb f (1+\alpha f) \br / \rc }
{\alpha (1-\alpha) (z-1) + z\beta \Vb \lambda (1+\alpha f)^2} \qquad (\Teff < \Tc),
\end{equation}

Using the periodicity condition $\rho(0) = \rho(1)$ then allows to express the condensate width as function of the stationary current $\bJ$.
Namely, we find after some algebra that
\begin{equation}
w = (1-\alpha)\left(1 - \frac{\ln(\chi)}{\beta \Vb f}\right) , \qquad \chi \equiv \frac{\br \beta \Vb f(1+\alpha f)(z + \alpha f) - \rc\left[(1-\alpha)(z-1) + \beta \Vb \lambda (1+\alpha f)^2 \right]}
{\br z \beta \Vb (1+\alpha f)^2 + \rc\left[(1-\alpha)(z-1) - \beta \Vb z \lambda (1+\alpha f)^2 \right]} .
\end{equation}
Although the expression of $w$ is fairly complicated, it reduces close to the transition and for vanishing temperatures to
\begin{equation}
w \underset{\Teff \lesssim \Tc}{\propto} \frac{\Tc}{\Teff} - 1, \qquad w \underset{\Teff \ll \Tc}{\simeq} (1-\alpha)\left(1 + \frac{\kT}{\Vb f}\ln\left[(1+\alpha f)(1-\lambda\rc/\br)\right]\right),
\end{equation}
where the first expression is given in terms of a complicated prefactor which is unimportant to specify here.
Similarly, the condensate fraction which can be calculated from
\begin{equation}
\phic = 1 - \frac{1}{\br}\left( \int_0^\alpha \rmd x \, \rho(x) + \int_{x{^*}}^1 \rmd x \, \rho(x) \right),
\end{equation}
is a complicated expression of limited practical use. 
Nevertheless, it is relatively straightforward to derive the following limiting behaviours
\begin{equation}
\phic \underset{\Teff \lesssim \Tc}{\propto} 1 - \frac{\Teff}{\Tc} , \qquad 
\phic \underset{\Teff \to 0}{\sim} 1 - \frac{\alpha^2 f}{(1+\alpha f)\lambda} \qquad (f > 0).
\end{equation}

\subsection{The Thermodynamics of condensation}

The average internal energy is defined as $\langle U\rangle = \int_0^1 U(x)\rho(x)\,\mathrm{d}x$. Above the transition threshold $\Tc$, the average energy per particle can be rewritten as
\begin{equation}
    \label{eq:internal_energy_fluid}
    \frac{\langle U\rangle}{N} = \frac{\rho_0}{N}\int_0^1 U(x)R(x)\mathrm{d}x  \qquad (f < 0, \, \Teff > \Tc)\,,
\end{equation}
where $N$ is the total number of particles ($N = \br$ in the chosen units).
In the subcritical condensed regime, the condensate energy depends only on the external driving force, not on the temperature. The energy per particle is then
\begin{equation}
    \label{eq:internal_energy_condensed_subcritical}
    \frac{\langle U\rangle}{N} = \frac{\rc}{\br} \int_0^1 U(x) R(x)\mathrm{d}x - \alpha F \phic\, \qquad (f < 0, \, \Teff \le \Tc)\,.
\end{equation}

For supercritical condensed systems, however, the expression for the internal energy is more complicated. Splitting the integral in three parts, corresponding to the three intervals $[0;\alpha]$, $[\alpha;x^*]$ and $[x^*;1]$ with $x^* \equiv \alpha + w$, yields
\begin{equation}
    \label{eq:internal_energy_condensed_supercritical}
    \frac{\langle U\rangle}{N} = \frac{\rc}{N}\int_0^\alpha U(x) R(x)\mathrm{d}x + \frac{\phic}{w}\int_\alpha^{x^*} U(x)\mathrm{d}x + \frac{\rc}{N}\int_{x^*}^1 U(x)R(x)\mathrm{d}x\, \qquad (f>0)\,.
\end{equation}
The second integral corresponds to the energy that is contained in the extended condensate, and it can be simplified as
\begin{equation}
    \frac{\phic}{w}\int_\alpha^{x^*} U(x)\mathrm{d}x = -\phic \Fc \left[\alpha + f\left(\alpha + \frac{w}{2}\right)\right]\,,
\end{equation}
such that the average energy is
\begin{equation}
    \label{eq:internal_energy_condensed_supercritical_2}
    \frac{\langle U\rangle}{N} = \frac{\rc}{N}\left[\int_0^\alpha U(x)R(x)\mathrm{d}x + \int_{x^*}^1 U(x)R(x)\mathrm{d}x\right]-\phic \Fc \left[\alpha + f\left(\alpha + \frac{w}{2}\right)\right]\,.
\end{equation}
The integrations occurring in equations \eqref{eq:internal_energy_fluid} to \eqref{eq:internal_energy_condensed_supercritical_2} can be performed exactly, but they result in complicated expressions that are of limited practical value. However, for subcritical strong and weak confinement, respectively, the energy becomes
\begin{align}
    \frac{\langle U\rangle}{N \Vb}&\underset{\beta \Vb \gg 1}{\sim} -\frac{f}{2 (1-\alpha)}\begin{cases}
        1 & \Teff>\Tc\\
        2\alpha + (1-2\alpha)\frac{\rc}{\br} & \Teff\leq \Tc
    \end{cases}\label{eq:energy_strong}\\
    \frac{\langle U\rangle}{N\Vb}&\underset{\beta \Vb \ll 1}{\sim}\frac{1}{2}\begin{cases}
        1-\frac{1+f}{1-\alpha} & \Teff>\Tc\\
        \frac{\rc}{\br}-\frac{1+f}{1-\alpha} \left[2\alpha + \frac{\rc}{\br}(1-2\alpha)\right]\,. & \Teff\leq \Tc
    \end{cases}
\end{align}
For weak confinement, the internal energy corresponds exactly to the results found in ref.~\cite{Mahault_2020} if the rescaled force is set to $f=-1$. Driving the system hence results in a linear decrease in energy as a function of the driving force. For strong confinement, however, the energy shows a clear departure from the ideal Bose gas, since is independent of $\Teff$.

The heat capacity $ C = \mathrm{d}\langle U\rangle/\mathrm{d}\Teff$ can be rewritten as $ C = -k_{\rm B}\beta^2\,\mathrm{d}\langle U\rangle/\mathrm{d}\beta$, and in the weak confinement limit it scales as
\begin{align}
    \frac{C}{N k_{\rm B}} &\underset{\beta \Vb \ll 1}{\sim}\frac{(\beta \Vb)^2}{12} \left[1-\frac{1-2\alpha}{1-\alpha} (1+f)\right]\begin{cases}
        1 & \Teff>\Tc\\
        4\, \frac{\rc}{\br} & \Teff\leq \Tc\,,
    \end{cases}
\end{align}
which again reduces to the known result for the undriven $f=-1$ case. In the strong confinement limit, the exact expressions become too complicated to study the asymptotic behaviour. However, numerical analysis reveals that $C/Nk_{\rm B} \sim {\rm const}$ for $\Teff>\Tc$ and $C/Nk_{\rm B}\sim 1/\beta \Vb$ for $\Teff\leq \Tc$, similar to the ideal Bose gas results of ref.~\cite{Golestanian_2019}.

\section{Details about numerical simulations}

Our numerical simulations of equation (1) of the main text were performed using the finite volume scheme proposed in ref.~\cite{Filbet_2012}, which is tailored to handle the steep solutions associated with the condensed phase.
The main idea of this scheme is to merge the convective and diffusive parts of the current, which we thus write as
\begin{equation}
    J(x,t) = - g(\rho(x,t)) \partial_x \left[ U(x) + h(\rho(x,t))  \right] \equiv g(\rho(x,t)) \partial_x A(x,t),
\end{equation}
where we have defined 
$g(\rho) \equiv M(\rho) \rho$ and assumed $M(\rho)>0$ such that we can define a function $h(\rho)$ satisfying
\begin{equation*}
    \partial_x h(\rho) \equiv \frac{D(\rho)}{M(\rho)}\partial_x \ln(\rho).
\end{equation*}

Following the finite volume approach in ref.~\cite{leveque_2002}, we discretize the spatial domain into cells of width $\Delta x$ and define for a given function $\zeta(x,t)$ its average over the $i^{\rm th}$ cell as $\zeta_i(t)$.
The continuity equation is then expressed for the $i^{\rm th}$ cell as
\begin{equation} \label{eq_contiuity_disc}
    \frac{\rmd\rho_i}{\rmd t} = - \frac{J_{i+1/2} - J_{i - 1/2}}{\Delta x}\,,
\end{equation}
where $J_{i \pm 1/2}$ denote the numerical approximations of the current at the boundaries of the cell, which are defined as
\begin{align*} \label{eq:numericalflux}
    J_{i + 1/2} & = \frac{A_{i+1/2}}{2} \left( g(\rho_i) + g(\rho_{i+1}) \right) - \frac{|A_{i + 1/2}| \alpha_{i + 1/2}}{2} \left( \rho_{i + 1} - \rho_{i} \right) , \\
    A_{i + 1/2} & = -\frac{U(r_{i+1}) - U(r_i)}{\Delta x} -\frac{h(\rho_{i+1}) - h(\rho_i)}{\Delta x} ,\\
    \alpha_{i + 1/2} & = \max(|g'(\rho)|) \qquad {\rm for} \quad \rho \in \{\rho_{i},\rho_{i+1}\} .\\
\end{align*}
The expression of $J_{i + 1/2}$ essentially corresponds to a local Lax–Friedrichs method~\cite{leveque_2002}, while the coefficient $\alpha_{i + 1/2}$ acts as a slope limiter that prevents numerical oscillations and ensures convergence to steady state.

Time integration of eq.~\eqref{eq_contiuity_disc} was performed via an explicit Euler scheme. For all data shown in the main text, we used space resolution $\rmd x = 0.01$ while the time step $\rmd t$ was taken between $10^{-4}$ and $10^{-5}$.
The diffusivity edge threshold $\rc$ was moreover set to $\rc = 10$ in all simulations.

\end{document}